\documentclass[conference]{IEEEtran}
\IEEEoverridecommandlockouts
\usepackage[table]{xcolor}
\usepackage{tikz}
\usepackage{url}
\usepackage{cite}
\usepackage{amsmath,amssymb,amsfonts}
\usepackage{algorithmic}
\usepackage{graphicx}
\usepackage{makecell}
\usepackage{booktabs}
\usepackage{comment}

\def\BibTeX{{\rm B\kern-.05em{\sc i\kern-.025em b}\kern-.08em
    T\kern-.1667em\lower.7ex\hbox{E}\kern-.125emX}}

\makeatletter
\newcommand{\linebreakand}{%
  \end{@IEEEauthorhalign}
  \hfill\mbox{}\par
  \mbox{}\hfill\begin{@IEEEauthorhalign}
}
\makeatother
    
\begin{document}

\title{A Herd of Young Mastodonts: \\ the User-Centered Footprints of Newcomers \\ After Twitter Acquisition
}

\author{\IEEEauthorblockN{Francesco Di Cursi}
\IEEEauthorblockA{\textit{Istituto di Informatica e Telematica} \\
\textit{Consiglio Nazionale delle Richerche}\\
Pisa, Italy \\
francesco.dicursi@iit.cnr.it}
\and
\IEEEauthorblockN{Chiara Boldrini}
\IEEEauthorblockA{\textit{Istituto di Informatica e Telematica} \\
\textit{Consiglio Nazionale delle Richerche}\\
Pisa, Italy \\
chiara.boldrini@iit.cnr.it}
\linebreakand
\IEEEauthorblockN{Andrea Passarella}
\IEEEauthorblockA{\textit{Istituto di Informatica e Telematica} \\
\textit{Consiglio Nazionale delle Richerche}\\
Pisa, Italy \\
andrea.passarella@iit.cnr.it}
\and
\IEEEauthorblockN{Marco Conti}
\IEEEauthorblockA{\textit{Istituto di Informatica e Telematica} \\
\textit{Consiglio Nazionale delle Richerche}\\
Pisa, Italy \\
marco.conti@iit.cnr.it}
}

\maketitle
\begin{tikzpicture}[remember picture,overlay]
\node[anchor=south,yshift=10pt] at (current page.south) {\fbox{\parbox{\dimexpr\textwidth-\fboxsep-\fboxrule\relax}{
  \footnotesize{
     \copyright 2024 IEEE. Personal use of this material is permitted.  Permission from IEEE must be obtained for all other uses, in any current or future media, including reprinting/republishing this material for advertising or promotional purposes, creating new collective works, for resale or redistribution to servers or lists, or reuse of any copyrighted component of this work in other works.
  }
}}};
\end{tikzpicture}

\begin{abstract}
The tremendous success of major Online Social Networks (OSNs) platforms has raised increasing concerns about negative phenomena, such as mass control, fake news, and echo chambers. In addition, the increasingly strict control over users' data by platform owners questions their trustworthiness as open interaction tools. These trends and, notably, the recent drastic change in X (formerly Twitter) policies and data accessibility through public APIs, have fuelled significant migration of users towards Fediverse platforms (primarily Mastodon). In this work, we provide an initial analysis of the microscopic properties of Mastodon users' social structures. Specifically, according to the Ego network model, we analyse interaction patterns between a large set of users (\emph{egos}) and the other users they interact with (\emph{alters}) to characterise the properties of those users' ego networks. As was observed previously in other OSNs, we found a quite regular structure compatible with the reference Dunbar's Ego Network model. Quite interestingly, our results show clear signs of ego network formation during the initial diffusion of a social networking tool, coherent with the recent surge of Mastodon activity. Therefore, our analysis motivates the use of Mastodon as an open ``big data microscope'' to characterise human social behaviour, making it a prime candidate to replace those OSN platforms that, unfortunately, cannot be used anymore for this purpose.


\end{abstract}

\section{Introduction}
\label{section:introduction}

Online Social Networks (OSNs) have become an essential part of modern life. This growth reflects a profound human need for connection and community. However, users are increasingly aware of the hidden costs of ``free'' services: their personal data, behavior, and preferences are the valuable commodity that fuels these platforms. The centralized nature of OSNs, often owned by private companies, raises concerns about data control and potential misuse. % Users both knowingly and unknowingly contribute vast amounts of data, which can be exploited for targeted marketing or, as the Cambridge Analytica scandal revealed, even for mass manipulation during sensitive periods ~\cite{hinds2020wouldn}.
The problems deriving from centralized Online Social Networks (e.g., privacy concerns and untrustworthy policies) are among the factors that lead users to explore different types of online social networks, such as distributed ones. Among these, the Fediverse is a network of interconnected social media platforms that use a common protocol, most notably ActivityPub, to communicate with each other. Unlike traditional social media platforms like Facebook or Twitter, the Fediverse is not controlled by a single company. It is made up of many independent servers (called ``instances'') run by individuals, organizations, or communities. The ActivityPub protocol allows these instances to communicate and share content. Most Fediverse software is free and open source: this fosters transparency and encourages innovation. Finally, the Fediverse generally does not useS algorithms to curate content, hence eliminating a direct source of algorithmic bias.
Mastodon\footnote{\url{https://joinmastodon.org/}} is a free and open-source microblogging platform, and it is arguably the most well-known member of the Fediverse. Similarly to Twitter/X, Mastodon allows users to post short messages (called ``toots'') and share images, videos, and links. %Differently from Twitter/X, Mastodon does not allow quoting (i.e. repost + comment). Another difference is the length of a toot: it can consist of up to 500 characters, although this may vary across instances. In Mastodon, the timeline is chronological, with no algorithm manipulation, allowing users to see posts in the order they were created.
Mastodon gained widespread popularity after the acquisition of Twitter/X by Elon Musk in 2022~\cite{mastodon_nyt,mastodon_guardian}. The user experience on Mastodon is very similar to Twitter/X and this led many users looking for an alternative to try the service. Additionally, Mastodon is particularly attractive for research due to its free API, which enables researchers to download and analyze large datasets with ease.

%Mastodon has attracted the attention of researchers since its inception. Many of the major works leveraging social network analysis use "follower/followee" relationships in order to understand how decentralization impacts on user behaviour and most importantly on federation, focusing on instances~\cite{la2022discovering,la2023polarization,la2022information,zignani2019footprints}. Other important delved aspects are mechanical and cognitive pressures toward centralization on such a decentralized architecture \cite{raman2019challenges} and the migration triggered by Musk's acquisition of Twitter \cite{he2023flocking}. Mastodon is especially appealing for research purposes due to its free API. 

\begin{figure}[ht]
\begin{center}
\includegraphics[scale=0.25]{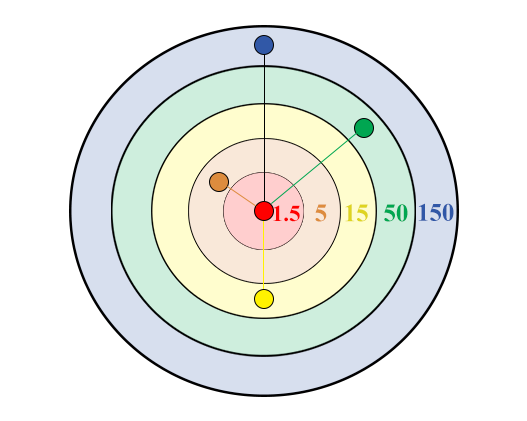}%\vspace{-5pt}
\caption{Layered structure of human ego networks}
\label{fig:egonet}%\vspace{-15pt}
\end{center}
\end{figure}

In this study, we explore social networks on Mastodon from a novel perspective—focusing on ego networks, a dimension that has not yet been examined on the platform.
Ego networks, a graph-based representation of an individual's social connections, are a common tool for studying interpersonal relationships across multiple interaction types, from offline exchange of messages, to phone calls, to online social networks~\cite{everett2005ego,lin2001social,mccarty2002structure,hill2003social}. These ego networks\footnote{This type of ego network is often referred to as Dunbar's ego network, in contrast to other ego networks that also include connections between alters.}, where the individual (ego) is central and connected to their peers (alters) through edges representing interaction frequency, are important because their structure is known to significantly influence social behaviors like resource sharing, collaboration, and information diffusion~\cite{sutcliffe2012relationships}. The ego-alter tie strength is typically computed as a function of the frequency of interactions between the ego and the alter. Grouping these ties by their strength, ego networks normally exhibit a layered structure (illustrated in Figure~\ref{fig:egonet}). Inner circles contain closer social connections, while outer circles represent more distant ones. The social brain hypothesis from evolutionary psychology~\cite{dunbar1998social} suggests that these layers, limited by the Dunbar number (the maximum number of \emph{meaningful} relationships an individual can maintain, around 150), reflect the brain's cognitive capacity for social relationships~\cite{hill2003social,Zhou2005}. Figure~\ref{fig:egonet} shows the typical sizes of each layer (1.5, 5, 15, 50, 150 alters). %Relationships beyond the Dunbar number are considered mere acquaintances with negligible impact on cognitive resources.
The grounding of this quantitative model on cognitive capacity arguments justifies why the very same structures (with approximately the same number of alters per layer) have been found across a wide range of social interaction means, from offline exchange of messages to very popular OSNs such as Facebook and Twitter/X~\cite{Dunbar2015}.

The contribution of this paper is twofold. We first analyse the recent surge of users and activity in Mastodon, in the period following the Twitter/X acquisition of 2022. Specifically, we notice a steep increase in the number of subscribed users, in the level of their activity (which is sustained over time), and in the level of direct interactions between users. \emph{These findings justify the use of Mastodon as a relevant platform for analysing human social behaviour}. Then, we set out to investigate whether Mastodon ego networks show structural properties compatible with Dunbar's ego network model. \emph{We find that the ego networks on Mastodon are compatible with those found on other major OSNs such as Twitter and Facebook at a similar evolution stage.} More specifically, they are comparable and compatible with the ``canonical'' Dunbar's model, in terms of the number of circles (with a preponderance of 4 and 5 layers) and scaling ratio ($\sim$3) across circles. We also observe, with respect to the Dunbar's model, a significantly lower number of alters, primarily in external layers. This has also been previously observed in Facebook and Twitter/X, at an evolution stage similar to that of Mastodon in this analysis. This is also compatible with the cognitive models grounding the Dunbar's model, as it is known that external layers need significant time in order to stabilise and become fully formed. Therefore, the findings presented in this paper indicate that Mastodon ego networks show similar patterns observed in many relevant social networking platforms, making it a quite promising candidate for studying more broadly human social relationships and resulting behaviours.

The paper is organised as follows. Section~\ref{section:preliminaries} contains a brief overview of the Fediverse and Mastodon.
Section~\ref{section:background} reviews relevant literature on Mastodon, with an emphasis on studies of its social graph.
Section~\ref{section:collection} details our data collection strategy using the Mastodon API and presents basic statistics on the retrieved data.
Section~\ref{section:methods} provides a description of ego networks and of the methods to construct them.
Section~\ref{section:results} offers an analysis of user activity in the collected dataset, both before and after acquisition. Finally, Section~\ref{section:structural_analysis} investigates the structural properties of the ego networks of Mastodon users, while Section~\ref{section:conclusion} summarizes the key findings of the study.

\section{A primer on Mastodon}
\label{section:preliminaries}

Mastodon, the central hub of the Fediverse, is a decentralized microblogging platform that resembles Twitter but with significant differences. The platform operates on two interconnected layers: instance-to-instance and user-to-user communication. The protocol linking these layers allows instances to function ``under the hood" when users from different instances interact, whether on the same or different services.

This architecture creates three distinct timelines that shape the user experience: the personal timeline (showing activity from followed users across instances), the local timeline (showing activity from users within the same instance), and the federated timeline (showing activity from any user in any instance where at least one local user is subscribed).

The federated timeline arises from the instance-to-instance relationship: when user $i$ from instance $A$ follows user $j$ from instance $B$, instance $A$ subscribes to instance $B$, exposing each instance's content to the other in their federated timelines. Instance admins can control access by making instances private, or by blacklisting certain users or problematic instances. Users also have control over their experience, with options to mute, report, or blacklist others.

Other notable differences include the absence of recommendation algorithms (posts, or ``toots" are displayed chronologically), and decentralized moderation, which occurs at the instance level based on local policies rather than universal censorship. Users can also customize their timelines by filtering out certain topics or users.
Mastodon's contact features are similar to Twitter’s, except for the absence of a ``quote'' feature (reposting with comments). Toots can be up to 500 characters long, though this may vary across instances. As with Twitter, communication on Mastodon can be direct (replies, mentions, or boosts, which function like retweets) or indirect (any toot that doesn’t engage another user). Users can also like and bookmark toots.

\section{Related works}
\label{section:background}

La Cava et al. conducted a series of studies on Mastodon using network science, focusing mainly on the instance level and using the following/followee feature to model relationships.
In the first study~\cite{la2022discovering}, La Cava et al. analyzed a directed network of instances derived considering the ties only between users in different instances, finding high clustering, reciprocal edges, degree disassortativity, sparseness, and small-world properties. They used the Louvain and Infomap algorithms and found that topic and language are key factors characterizing instances. Core decomposition revealed connections from core to periphery, reducing sectorization bias. Observations of the network of online instances at the time of data collection showed similar results but with a significant network shrinkage (-84\% instances and -65\% links), indicating structural stability.
Further analysis focused on user behavior w.r.t. content consumption in a decentralized architecture~\cite{la2022information}. By filtering out instances with fewer than 51 degrees and considering the top-5 instances, they identified the network backbone with properties similar to their previous findings~\cite{la2022discovering}. The core network consisted of \texttt{mastodon.social}, \texttt{pawoo.net}, \texttt{mastodon.xyz}, \texttt{mastdn.io}, and \texttt{octodon.social}. La Cava et al. found that users in larger instances tend to follow users on smaller ones. They also noticed few sink nodes (no followees, only followers) but many shell nodes (no followers, only followees), indicating boundary-spanning. Strong bridges (mutual relationships) and lurkers (identified through an eigenvector-based approach) were present, though few users acted as both.
In another study~\cite{la2023polarization}, they explored polarization using a signed network, identifying positive (following) and negative (instance ban) relationships. They found four groups: a neutral group (including non-Mastodon instances), two Mastodon-pure groups, and a ban-sink group (receiving negative links). The largest groups were the neutral and the pure groups, which showed a significant amount of positive interaction, while the ban-sink group contained inappropriate content and bot activity.

Zignani et al.~\cite{zignani2019footprints} studied the impact of decentralization on user behavior within Mastodon, finding that user behaviors vary significantly across instances. They noted that users are not strongly bound by their instances, with 35-40\% linked to other instances, but only 10\% connecting to instances in different countries, highlighting a strong linguistic-geographical assortativity. Users typically explore a few instances driven by topic and geography/language. Instances have diverse characteristics, with larger ones having higher local clustering coefficients. Degree distributions are heavy-tailed across all instances, with degree exponents ranging from 2.1 to 3. Their findings suggest that different instances have different footprints, which, in turn, condition user behavior.

Raman et al.~\cite{raman2019challenges} examined pressures toward centralization in a decentralized architecture by analyzing instances, toots, and both follower and federation graphs, comparing them to tweets. They identified at least three centralization pressures. \textbf{User-driven pressures}: an oligarchy of admins governs the Fediverse, with 5\% of instances hosting 90\% of users and 94\% of toots. Users in open instances are less active than those in closed ones, with single-user instances being the most active ones. \textbf{Infrastructure-driven pressures}: admins use the same cheap servers from major corporations like Google and Amazon, leading to high co-location of instances and the same underlying concerns about privacy. \textbf{Content-driven pressures}: few hashtags account for the majority of toots (i.e., tech, game and art) although few sensitive toots attract the majority of users (i.e., adult content). Few linguistically and culturally bounded instances dominate subscriptions in a strongly assortative manner (JP, USA, FR and GR). Finally, outages in 10 instances could remove 60\% of global toots and the authors suggest replication schemata (i.e., federation-based and random, with the latter outperforming the former given the higher probability of selecting instances on different hosting providers).

He et al.~\cite{he2023flocking} investigated the migration from Twitter to Mastodon following Musk's takeover. They found that 20\% of new Mastodon profiles were created before the decline of Twitter, with users favoring popular instances. Larger instances attract more users, but smaller ones have more active users. Migration is influenced mainly by peer pressure, with users following their followees to the same instances and the same phenomenon is observed in instance migration. Finally, platform migration is not definitive: the activity increases on Mastodon but does not decrease on Twitter, with most users mainly posting different or similar content on each platform and a few using cross-platform posting tools.

Existing studies on Mastodon have not approached relationships on the platform in the way this work intends to. Previous research has primarily focused on ``follower/followee" relationships, which represent a limited model of user-to-user interaction. While useful for examining platform engagement, this model does not fully capture genuine social interactions.
Notably, only one prior study has analysed ego networks in  Mastodon, but it employed a graph-teoretical definition and focused on triadic closures within and across instances. In contrast, this work employs Dunbar's ego network model to explore the structural properties essential for understanding social interactions. We assess the developmental stages of these networks and compare them with offline communication and centralized OSNs. By constructing an interaction graph based on directed communication, we extract and analyze ego networks, providing deeper insights into social dynamics on Mastodon.

\section{Data collection}
\label{section:collection}

The data collection started in January 2024 and ended in March 2024. December 31, 2023 is set as the end of the observation period for all the collected timelines: this means that only toots published before that date are considered. The collection (agnostic to Mastodon instances) was carried out with a snowball sampling strategy using the public Mastodon API, starting from a random user highly active in the first years of Mastodon (i.e., 2016/2017) and subscribed to the first and main Mastodon instance (i.e., \texttt{mastodon.social}). For building ego networks, we are interested in direct communications between users, and these are the types of interactions we collect.

Starting from the aforementioned user, the username, the instance and the user ID of all their alters are collected, prioritizing the active ones (i.e., those having at least 2 contacts happening at least once a year). Then, the first active alter is visited and the same procedure is applied. The procedure is repeated on each alter of each user, stopping at 2000 collected users. In this way, all collected users belong to the same connected component.
The Mastodon API is used to fetch all the toots of a user. By scrolling through the timeline and filtering for dates, the whole activity of a user can be retrieved. Table~\ref{tab:user_stats} reports some statistics about the collected users, where column \emph{ego} refers to the collected profiles, column \emph{alters} to the set of alters of the collected users and column \emph{active alters} refers to the set of alters contacted at least once a year by the collected egos. Table~\ref{tab:user_stats} shows that the directed (i.e., social) activity in the collected sample is significant. 

Moreover, as a by-product of the collection strategy, almost all collected users appear as alter for at least another user, with many of them also being active. The number of users labelled as bots (directly by Mastodon) is negligible. The resulting network is large and comprises approximately 132,000 nodes (i.e., users + alters) and $\sim$1M directed links, for a total of $\sim$3M user-to-alter interactions.

\begin{table*}%[t]
    % %\vspace{-10pt}
    \centering
    \caption{Summary statistics for the Mastodon users in our dataset}
    \label{tab:user_stats}
    %\scriptsize
    % \resizebox{\columnwidth}{!}{
    \begin{tabular}{@{}ccccccccc@{}}
    \toprule
        \textbf{ego} & \textbf{alters} & \textbf{\makecell{bot\\ego}} & \textbf{\makecell{bot\\alters}} &  \textbf{\makecell{active\\alters}}	& \textbf{\makecell{ego being\\ alters}} & \textbf{\makecell{ego being\\active alters}} &  \textbf{\makecell{directed\\links}} & \textbf{\makecell{all\\interactions}} \\
        \midrule
        1,999 & 130,492 &  19 & 1,762 & 71,272 & 1,849  & 1,672 & 973,217 & 2,932,049 \\
        \bottomrule
    \end{tabular}
    % }
    % %\vspace{-5pt}
\end{table*}

Table~\ref{tab:toots_types} reports the number of all toots according to their types. We split toots into two categories: (i) directed toots, comprising replies, mentions and boosts (i.e., repost), and (ii) undirected, i.e., any other toot.
The collected sample contains $\sim$4M toots, the vast majority directed ones ($\sim$2.6M toots\footnote{Note that the number of total interactions in Table~\ref{tab:user_stats} is slightly higher because a single toot can include interactions with different alters.}) and considerably less undirected ones ($\sim$1.3M).  Other columns account for communication features only, ignoring the repetition of a contact feature in a single toot in case of repeatable ones (e.g., a toot with multiple mentions is counted as 1 toot with mentions). Among all toots, $\sim$1M are replies and $\sim$1.6M are boosts. These 2 directed features account for almost half the whole toots (i.e., almost the totality of directed toots). This further validates the collected sample for the analysis of direct communication. 

\begin{table*}[t!]
    % %\vspace{-10pt}
    \centering
    \caption{Summary statistics for the toots in our dataset} 
    % %\vspace{-40pt}
    \label{tab:toots_types}
    \scriptsize
    \begin{tabular}{@{}cccccccccc@{}}
    \toprule
    \textbf{toots} & \textbf{\makecell{dir\\toots}} & \textbf{\makecell{undir\\toots}} & \textbf{\makecell{plaintxt\\(no dir)}} & \textbf{\makecell{replies}} & \textbf{\makecell{toots w/\\hashtags}} & \textbf{\makecell{toots w/\\mentions}} & \textbf{\makecell{toots w/\\urls}} & \textbf{\makecell{boosts}} & \textbf{\makecell{toots w/\\multimedia}}\\ 
     \midrule
    3,965,007 & 2,626,141 & 1,321,860 & 464,902 & 1,043,683 & 391,851	& 208,874 & 697,266 & 1,663,551 & 357,083 \\
    \bottomrule
    % \begin{comment}\textbf{\makecell{toots w/\\polls}} &\end{comment} 
    % \begin{comment} 2,425 & \end{comment} 
    \end{tabular}
    % %\vspace{5pt}    
\end{table*}

\section{Methods}
\label{section:methods}
% %\vspace{-10pt}

\subsection{The ego network model: an overview}
\label{subsection: egonet_overview}
%\vspace{-5pt}

A popular approach to model relationships is to consider the immediate social surroundings of users. This approach has been developed in the anthropological literature for offline communications. A positive correlation between the size of the neocortex (particularly the prefrontal cortex) and the size of the social circles among primates and humans has been proved to be not a mere coincidence. Instead, sociability (along with its consequences, such as rules, roles, and coordination) has been proposed as the factor that fosters the evolution of the brain and, in turn, also of language~\cite{dunbar1992neocortex,dunbar1992co,aiello1993neocortex}.
These studies, besides posing sociability as a key factor in evolution, also observe the structure and the properties of user-centered social structures~\cite{dunbar1998social}.
%although with some critics ~\cite{lindenfors2021dunbar,stout2018human}.
The \emph{ego network} is a structure built upon the following elements: ego (the observed individual), alters (set of individuals contacted by the ego), tie strength (contact frequency, typically per year), circles (nested groups from the bigger and less intimate outermost one to the smallest and more intimate innermost one).
The number of actively maintained relationships is found, from the related literature, to be in the proximity of 150, posing those outside this number as ``inactive alters'' (i.e., casual relationships). The set of circles that contain these 150 active alters is the ego network, and the main properties of the ego network found in offline communication are reported in Table~\ref{tab:egonet_offline}. 

\begin{table*}%[t!]
% %\vspace{-5pt}
    \centering
    \caption{Ego network properties in offline ego networks}
    \label{tab:egonet_offline}
    %\scriptsize
    % \resizebox{\columnwidth}{!}{
    \begin{tabular}{ c  c  c  c  c  c }
    \toprule
       \makecell{\textbf{Layer}\\\textbf{number}} & \makecell{\textbf{Layer}\\\textbf{name}} & \textbf{Description} & \textbf{Alters} & \makecell{\textbf{Scaling ratio}\\\textbf{(alters)}} & \makecell{\textbf{Contact}\\\textbf{frequency}}   \\ \midrule
         0 & Super-support clique & partner/best friend &1.5 &  & once every 5 days \\ %\cline{1-4}\cline{6-6}
         1 & Support clique & close family & 5 &  & once every week \\ %\cline{1-4}\cline{6-6}
         2 & Sympathy group & close friends & 15 & $\sim 3$ & once a month \\ %\cline{1-4}\cline{6-6}
         3 & Affinity group & \makecell{friends/extended family/\\colleagues} & 50 &  & once every 6 months \\ %\cline{1-4}\cline{6-6}
         4 & Active network & meaningful relationships & 150 &  & once a year\\ \bottomrule
    \end{tabular}
    % }
    % %\vspace{5pt}
    %\vspace{-10pt}
\end{table*}

\subsection{How to extract ego networks}
\label{subsection:freq_and_layers}

Constructing an ego network entails extracting, for each user serving as ego, the list of its alters with the corresponding frequency of interaction. Then, a clustering algorithm is run on the interaction frequencies to extract the groups of alters, i.e., the social circles. Here, we follow the standard pipeline of the related literature on ego networks~\cite{arnaboldi2015online,boldrini2018twitter,toprak2022harnessing}.
As we explain in Section~\ref{section:collection}, our dataset (and in general social network datasets from which ego networks can be extracted) contains the social (i.e., directed) interactions of Mastodon users with other Mastodon users. Specifically, each user for which we have the timeline is an ego in our analysis. Thus, for each other user (i.e., alter) with which the ego interacts, we need to compute the contact frequency. Then, we filter out the alters contacted less frequently than once per year, the threshold for the active part of the ego network discussed above. For alters contacted just once, it is not possible to reliably compute the frequency, hence they are discarded. In addition, only relationships that have lasted at least six months are retained for our analysis, similarly to the related literature~\cite{arnaboldi2015online}.
%The collected data contains all the toots produced by users. The focus for ego network analysis is on those users starting their activity from the end of the acquisition onward. As already pointed in Section~\ref{subsection: egonet_overview}, the ego network accounts for active alters only. In order to proceed, among all toots, the directed ones have been retrieved by extracting those with reply tags, mention tags and reblogs. Then, a filtering for meaningful relationships is applied as follow. 
In other words, given a user $U_i$, its alter $U_j$, the end date of observation $T_{end}$ (i.e. 2023-12-31), and ego-alter contacts $C_{ij}$ starting at $T0_{ij}$ with annual frequency $F_{ij}$, all $C_{ij}$ are retained only if the following conditions hold: $T0_{ij} < T_{end} - \textrm{6 months}$,  $C_{ij}\geq2$ and $F_{ij}>1$.
$F_{ij}$ are then used as input for the Meanshift algorithm (one of the algorithms commonly used in the recent related literature~\cite{boldrini2018twitter,toprak2022harnessing}) in order to find the number of social circles that make up their ego networks. Meanshift~\cite{comaniciu2002mean} is a non-parametric clustering algorithm that iteratively shifts each data point towards the mode (the region of highest density) of its local neighborhood until convergence. Its bandwidth parameter controls the size of neighborhoods, and it can be automatically optimized with most off-the-shelf data analysis libraries (e.g., \texttt{scikit-learn}). This means that Meanshift automatically selects the optimal number of clusters (i.e., social circles) into which frequencies can be grouped, hence the layered ego network structure is not forced but, if present, emerges naturally from this grouping. 

%\vspace{-10pt}
\section{Validating Mastodon for ego network analysis} \label{section:results}
%\vspace{-5pt}

The existence of an ego network structure in online interactions requires a certain amount of cognitive effort from users on the considered platform. When social activity is low, ego network structures are not expected to emerge. Since these structures capture cognitive limits to social interactions, these limits do not appear when engagement is very low. For this reason, datasets used for ego network analysis must guarantee a certain amount of social activity to be suitable for such investigation. In this section, we address this aspect and investigate whether our Mastodon dataset meets this criterion. 

%\vspace{-10pt}
\subsection{Analysis of the overall daily activity}
\label{section:overall_activity}

\begin{figure*}[t!]
    %\vspace{-30pt}
    \centering
    \begin{minipage}[t]{.5\textwidth}
        \centering
        \includegraphics[width=\textwidth]{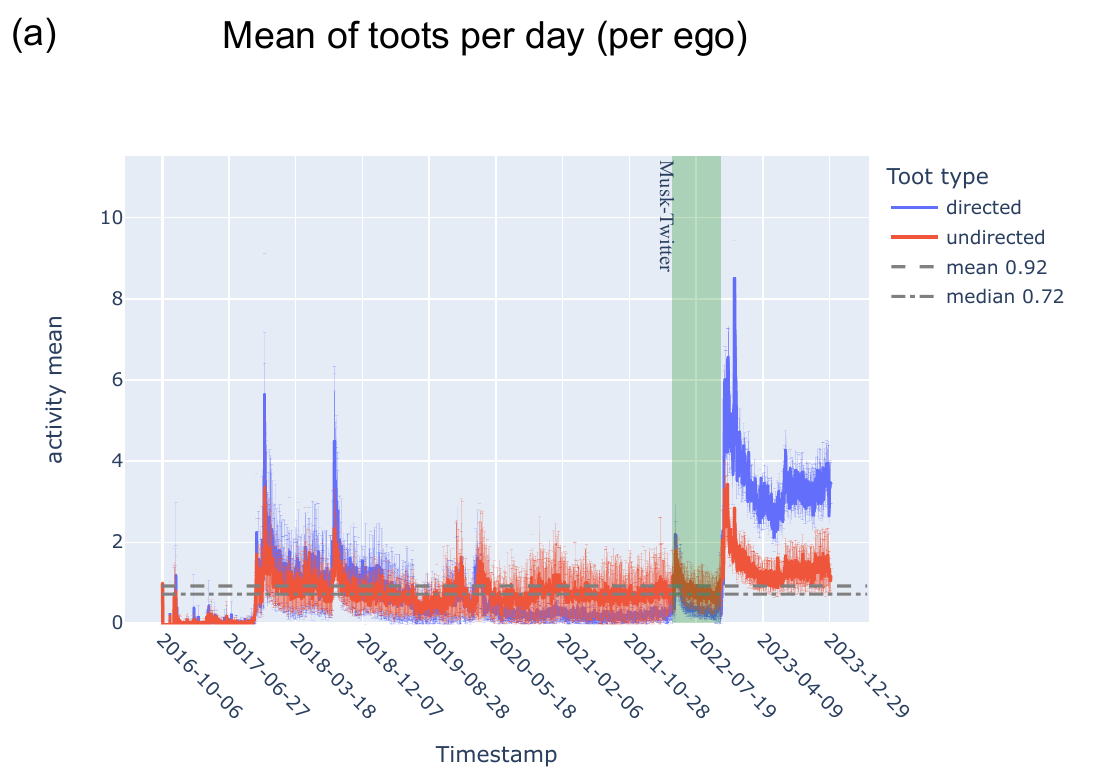}
        %\caption{}
        \label{subfig:all_timeline_only}
    \end{minipage}%
    \begin{minipage}[t]{.5\textwidth}
        \centering
        \includegraphics[width=\textwidth]{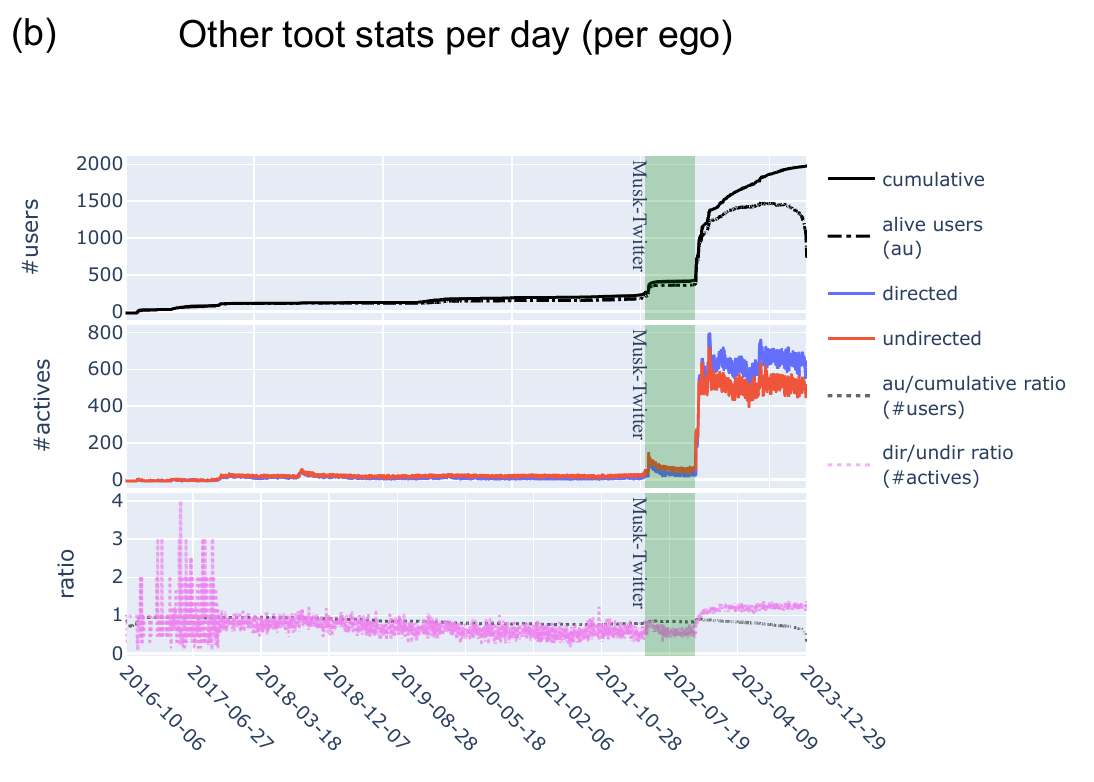}
        %\caption{}
        \label{subfig:all_timeline_other_info}
    \end{minipage}
    %\vspace{-10pt}
    \caption{Overview of the user activity over time. On the left (a), the average daily activity (measured as the average number of daily toots) per user. On the right (b), we show: the number of users that have posted once by date $x$ vs the number of users (which we call \emph{active}) that will continue posting after date $x$ (b, top panel), number of users engaging in directed vs undirected toots for each day (b, middle panel), ratios between the metrics in the top and middle plots (b, bottom panel).}
    \label{fig:all_timeline}
\end{figure*}

Figure~\ref{fig:all_timeline}(a) shows the average number of toots per day of the Mastodon users in our dataset. After Musk's acquisition of Twitter, we observe a substantial increase in the number of daily toots. 
In terms of the absolute number of users on the platform, the top panel in Figure~\ref{fig:all_timeline}(b) reveals a similar boost, with the user base (``cumulative'' curve, measuring the number of users having posted at least one toot by that date) more than quadrupling after the acquisition. However, an examination of the ``alive'' users on a given day (i.e., those who continue to post toots beyond that day, represented by the dot-dashed curve) indicates that the growth of users is not closely followed by that of alive users. This implies that some users tend to stop using the platform after they initially joined (e.g., they write a few tweets and then they stop). Despite this, alive users still grow by 3x with respect to the period before the acquisition.
The middle panel of Figure~\ref{fig:all_timeline}(b) shows daily alive users producing at least one toot (directed, in the blue curve, undirected, for the red curve) in that day. Again, there is a significant increase after the acquisition and a consistent correlation between directed and undirected toots. After the acquisition, the number of daily active users producing social toots stabilizes around 500, supporting the notion of a new, stable user base.
Finally, the bottom panel of Figure~\ref{fig:all_timeline}(b) compares the acquisition rate of new users and their communication preferences. The black dotted curve represents the ratio between active users and registered users over time. It indicates a steady retention rate with a slight decrease post acquisition due to many users trying and leaving the service. The pink curve indicates the ratio between directed and undirected toots, and it shows a shift from undirected to directed communication after the acquisition. 
Overall, the results in Figure~\ref{fig:all_timeline} suggest that Mastodon has seen a boost in active users and activities following Musk's acquisition of Twitter. As expected, not all new users are retained by the platform, but a stable user base of active users has clearly emerged.

Given the results in Figure~\ref{fig:all_timeline}, it is reasonable to divide the users into three categories: Aficionados (active both before and after the acquisition), Others1 (active only before), Others2 (active only after).
The 77\% of the collected users started their activity after the acquisition (see Table~\ref{tab:collected_types}). Given the divergence between these types (both quantitatively and qualitatively), analyzing these users together may lead to inconsistencies. Given that Others2 drastically outnumbers the other groups and that they are active in the period where the majority of directed communications take place, for this study, we decided to focus exclusively on the Other2 group. 

\begin{figure}
    \centering
    \includegraphics[width=\linewidth]{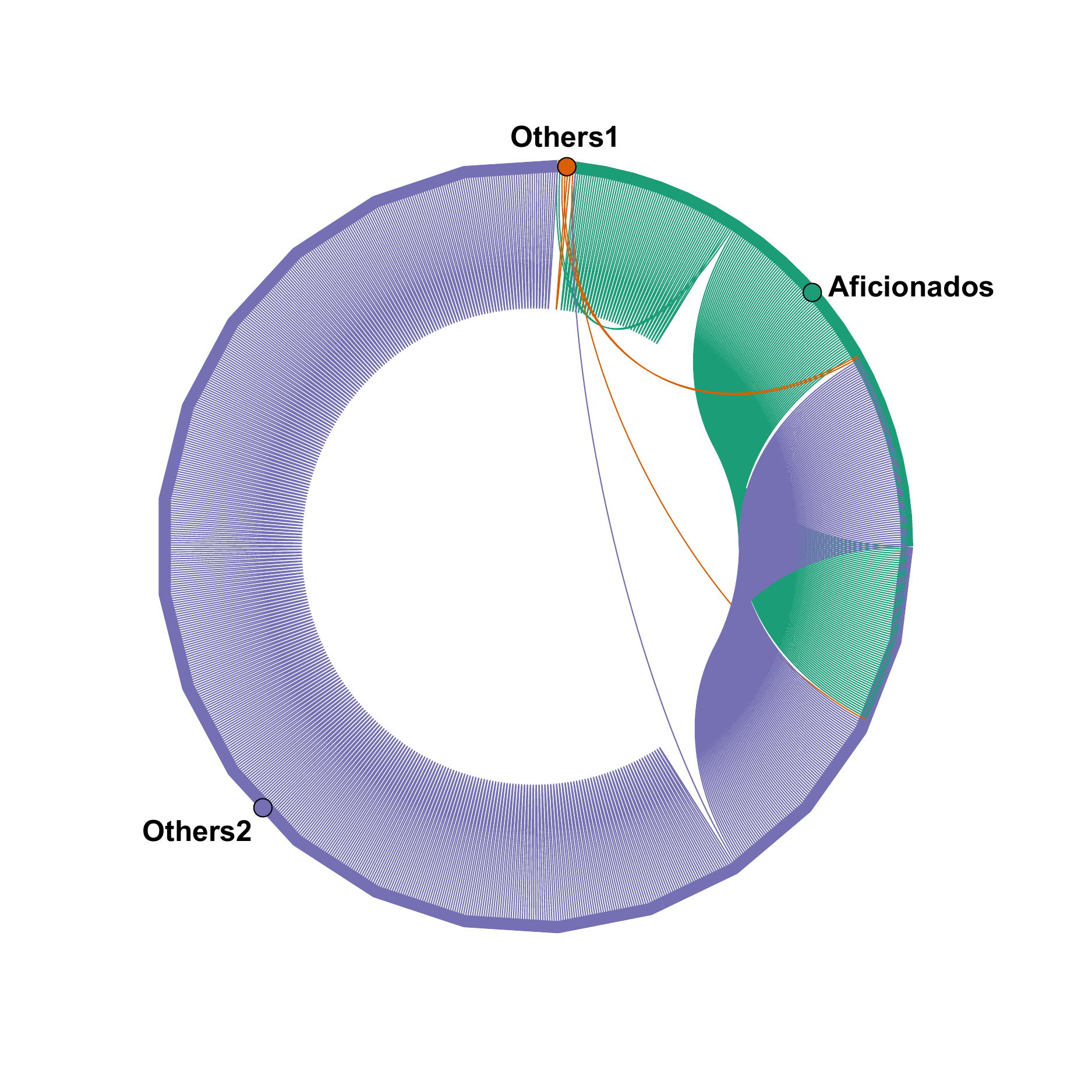}\vspace{-20pt}
    \caption{Interactions between different categories of users.}
    \label{fig:inter_vs_intra}
\end{figure}

We also analyzed whether intra-type interactions (e.g., Others2 to Others2) are more common than inter-type interactions, or vice versa. As shown in Figure~\ref{fig:inter_vs_intra}, Others2 tend to interact more with Others2, while Aficionados and Others1 are more equally divided between intra- and inter-type interactions. 

\begin{table}[t!]
    \caption{Types of collected users}
    \label{tab:collected_types}
    \centering
    \begin{tabular}{@{}ccc@{}}
    \toprule
       \textbf{type} & \textbf{count} & \textbf{\%}\\
       \midrule
        Aficionados & 404 & 20.21\%\\
        Others1 & 61 & 3.05\%\\
        Others2 & 1534 & 76.74\%\\
        \bottomrule
    \end{tabular}
    %\vspace{-10pt}
    \end{table}

\subsection{Analysis of the post-acquisition daily activity}
\label{section:post_acquisition_activity_analysis}

\begin{figure}[t!]
    %%\vspace{-15pt}
    \centering
    \begin{minipage}[t]{\columnwidth}
        \centering
        \includegraphics[width=\columnwidth]{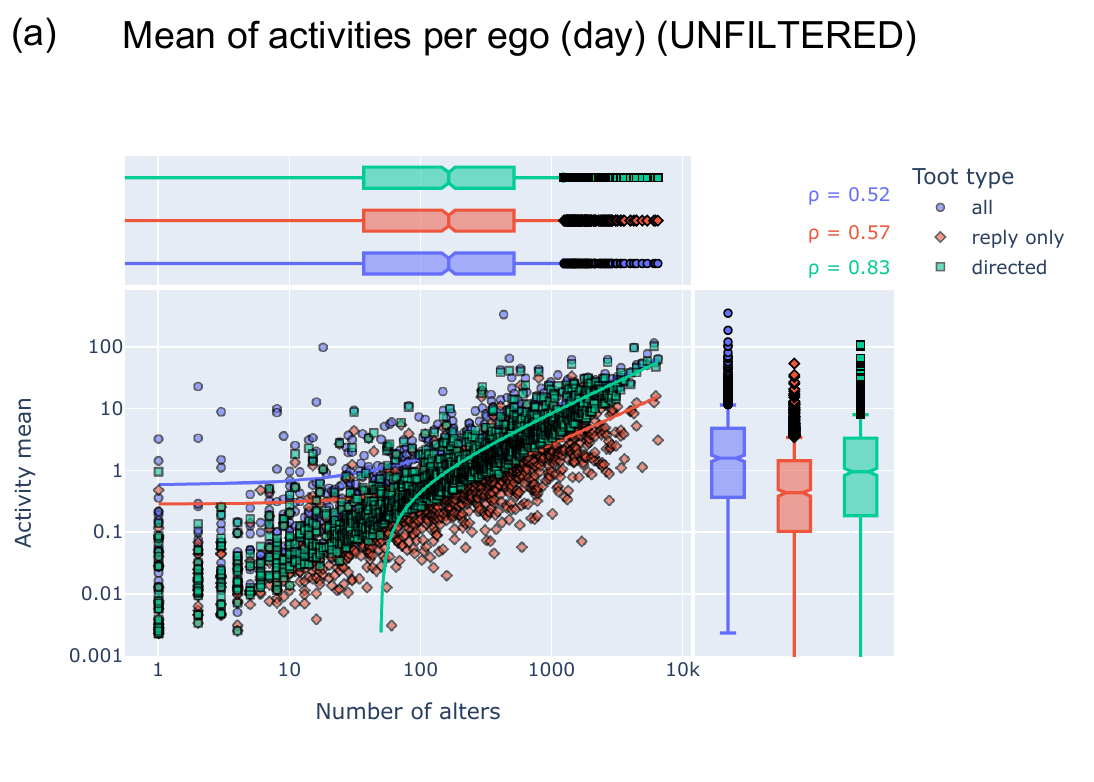}
        %\caption{All users and alters}
        \label{subfig:unfilt_ego_size_activity}
    \end{minipage}%
     
    \begin{minipage}[t]{\columnwidth}
        \centering
        \includegraphics[width=\columnwidth]{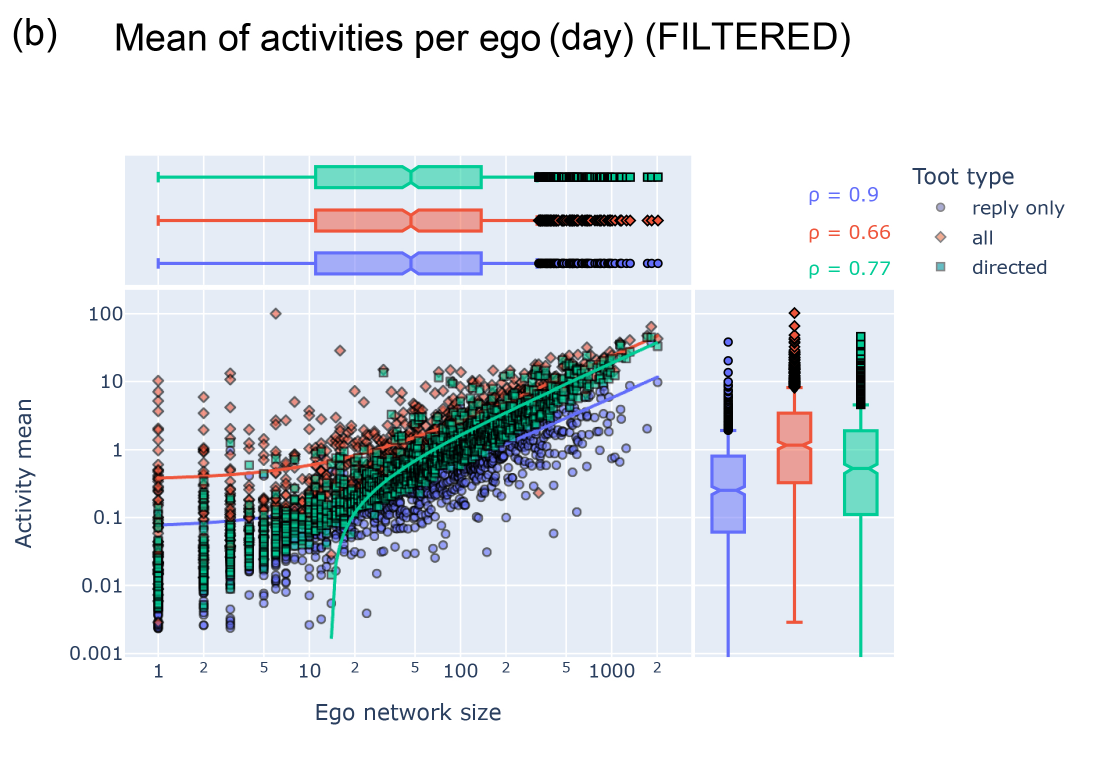}
        %\caption{Only active users and alters}
        \label{subfig:filt_ego_size_activity}
    \end{minipage}
    %\vspace{-30pt}
    \caption{User activity vs number of alters }
    \label{fig:ego_size_activity}
\end{figure}

Figure~\ref{fig:ego_size_activity} shows how the total number of alters per ego correlates with the average ego activity on Mastodon, both considering all egos and only those that pass the preprocessing/filtering steps described in Section~\ref{subsection:freq_and_layers}.
Each point corresponds to a user, where the $x$-coordinate represents its number of alters, and the $y$-coordinate represents the average number of toots posted per day by that user. 
Looking at the scatter plot and at the Pearson correlations (the $\rho$ values at the top right of the figures), in both cases, the activity is positively correlated with the number of alters, with higher values for the filtered data. This high positive correlation between the ego network size and their activity suggests that we are observing a growth phase. In other words, users are not saturated w.r.t. cognitive energies, and they can serve all of their alters with the same amount of time and attention. 
Let us now focus on the marginal distribution of the number of alters (box plots at the top). Unfiltered egos have a median three times higher than the filtered one ($\sim$150 and $\sim$50 respectively in Figures~\ref{subfig:unfilt_ego_size_activity} and~\ref{subfig:filt_ego_size_activity}), meaning that the ego networks when considering only filtered data are relatively small, possibly due to the short observation window (i.e., 1 year and a half). This is further explained in Section~\ref{section:structural_analysis} on the basis of Dunbar's models.
Looking at the marginal distribution of the average activity (box plots on the right), considering the minimal difference between ``all" and  ``directed" toots, the directed toots can be considered a good proxy of the overall activity on Mastodon, at least in the collected data. The difference between ``directed'' and ``reply only'', instead, suggests that there are many boosts and mentions.

\begin{figure}[t!]
    %\vspace{-5pt}
    \centering
    \includegraphics[width=\columnwidth]{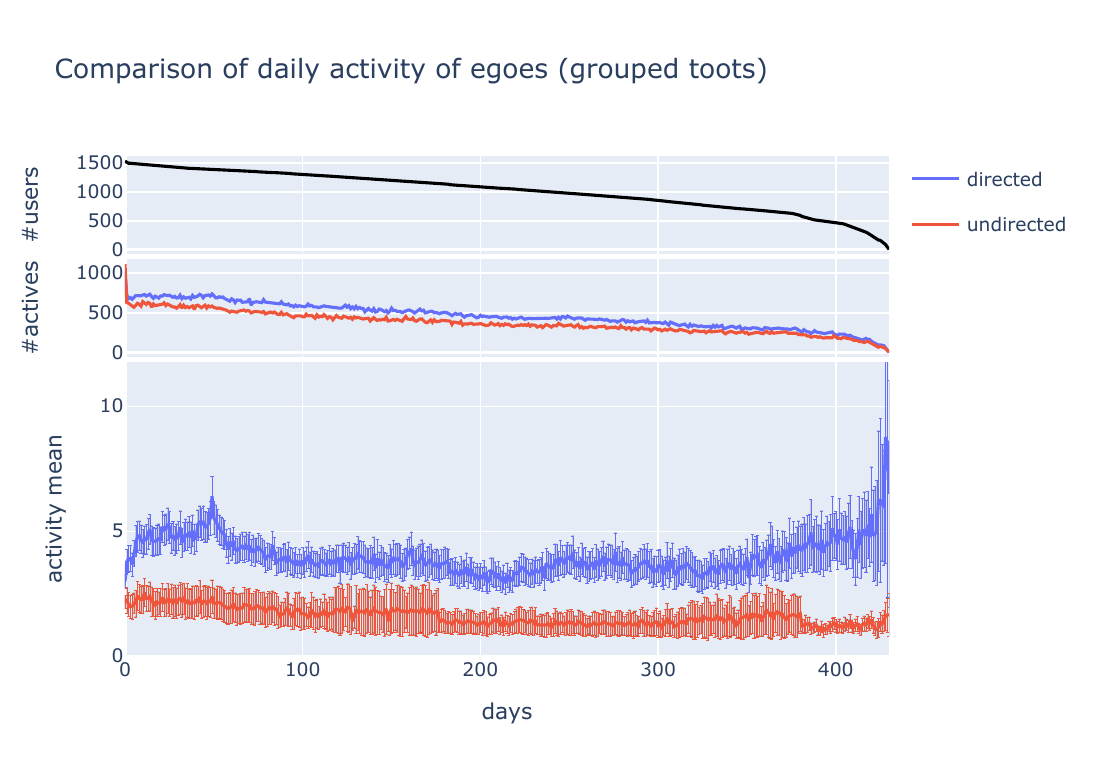}
    %\vspace{-20pt}
    \caption{Activity post-acquisition. On the $x$-axis, the lifespan of users (defined as the number of days since the first toot). On the $y$-axes, the plot shows: the number of users that have reached that lifespan (top), the number of users that have reached that lifespan and have continued posting directed/undirected toots (middle), the average number of daily toots split per type (bottom).}
    %\vspace{-20pt}
    \label{fig:others2_lifespan_comparison}
\end{figure}

Figure~\ref{fig:others2_lifespan_comparison} shows user activity over time on Mastodon for users in Others2. Their lifespans, from their first toot to December 31, 2023, are aligned for comparison.
The top panel (i.e., \#users) depicts the number of users who posted their first toot $x$ days before. While more than half of our $\sim$1500 users in the Others2 group have a lifespan of at least 300 days, only a small fraction of them reaches more than 400 days of lifespan.
The middle panel shows the number of users who generate at least one toot at least $x$ days after their first toot (in the post-acquisition phase). Note that while the top panel characterises the lifespan of users, whether they remain active or not, the middle panel focuses only on active users, i.e., those who continue posting toots. On a total of 1500 users, only 500 users remain regularly active. We observe a sudden drop of activity after the first days, meaning that several users suddenly reduce their commitment to the platform and their activity become more sporadic.
The third panel reveals the average toot production per day, with higher activity at the beginning (according to the well-known novelty effect of social media platforms) and at the end (the latter can be attributed to the survivorship bias, as long-term users tend to be more active than the general Mastodon population). Direct communication is initially high, declines slightly after 50 days, but then increases around 200 days, peaking at 400 days, indicating highly active stable users. Undirected communication is consistently lower, with notable declines around 200 and 400 days.
Overall, the data suggests an initial burst of activity driven by the novelty of the platform, most likely for users flocking to Mastodon after the Twitter acquisition, followed by more sporadic engagement and a stable, albeit small, group of users favoring direct communication for social interaction rather than undirected information diffusion.

\section{Ego networks of Mastodon users}
\label{section:structural_analysis}

In the previous section we have shown that our Mastodon dataset contains a group of stable and active users that can be leveraged for the ego network analysis. In the following, then, we discuss the properties of the ego networks of these Mastodon users. 
Recall that the method for building the ego networks and the necessary preprocessing steps can be found in Section~\ref{subsection:freq_and_layers}.

\begin{figure}[t!]
    %%\vspace{-15pt}
    \centering
    \begin{minipage}[t]{\columnwidth}
        \centering
        \includegraphics[width=\columnwidth]{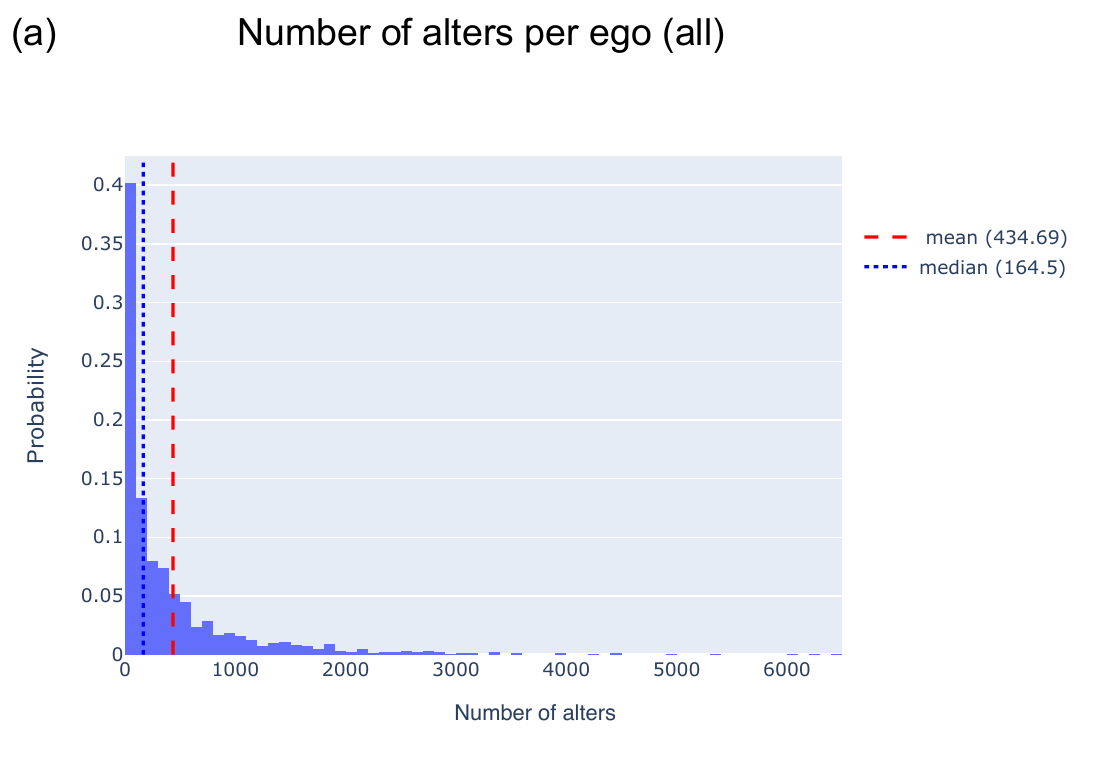}
        %\caption{All alters}
        % \label{subfig:n_alters_all}
    \end{minipage}%
    
    \begin{minipage}[t]{\columnwidth}
        \centering
        \includegraphics[width=\columnwidth]{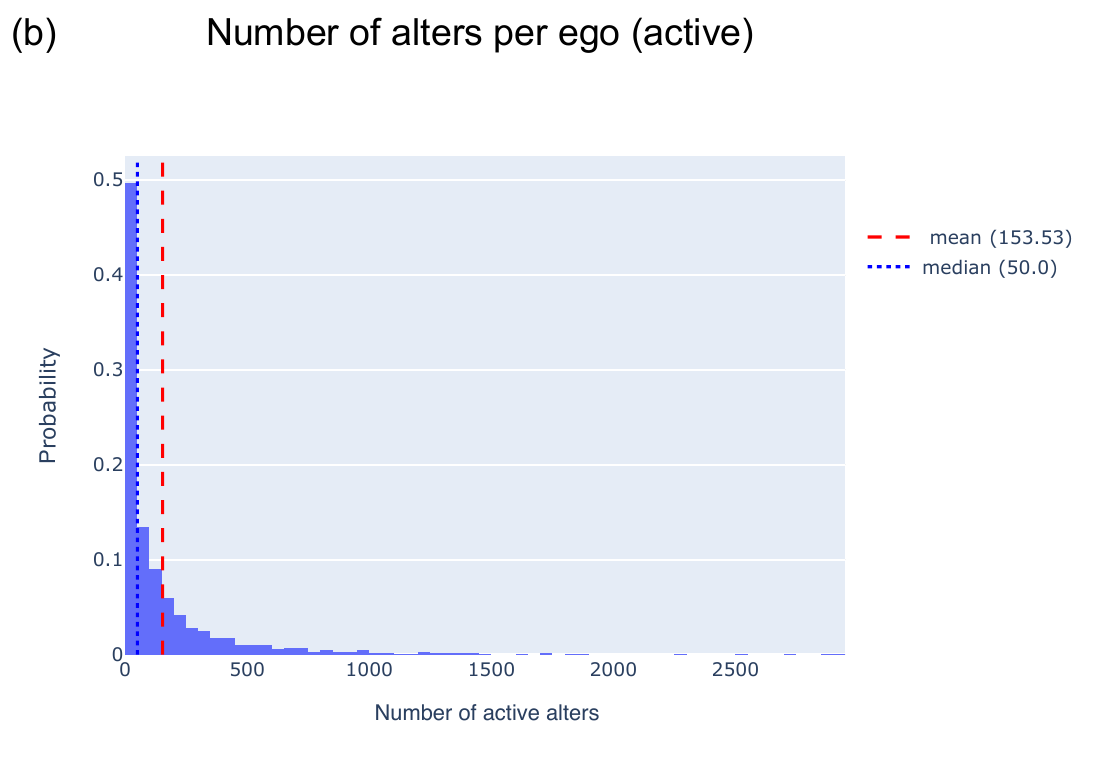}
        %\caption{Only active alters}
        % \label{subfig:n_alters_active}
    \end{minipage}
    %\vspace{-30pt}
    \caption{Number of alters per ego: (a) all alters, and (b) only active alters}
    %\vspace{-10pt}
    \label{fig:n_alters}
\end{figure}

Figure~\ref{fig:n_alters}(a) shows the average number of alters per ego. With respect to the ego network model, this is the size of the \emph{entire} ego network, not considering the difference between active and inactive alters (respectively, those contacted at least once vs less than once per year, see Section~\ref{section:methods}). Therefore, its large size (around 450) is not surprising and compatible with a layer called ``mega-band'' in the anthropology literature. However, it is more important to analyse the \emph{active} part of the ego networks only, shown in Figure~\ref{fig:n_alters}(b), as alters beyond this limit do not ``consume'' the cognitive resources of the ego.
The average number of active users (i.e., contacted at least once a year) is close to Dunbar's number ($\sim$150), while the inactive users are almost double.

\begin{figure}[t!]
    \centering
    \includegraphics[width=\columnwidth]{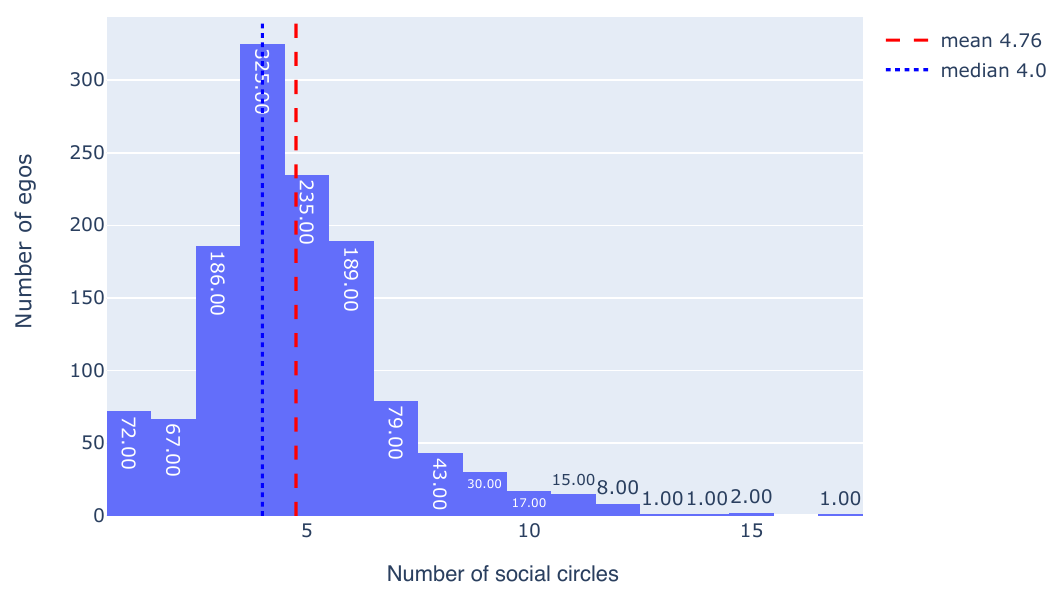}
    %\vspace{-10pt}
    \caption{Number of social circles in the Mastodon ego networks}
    %%\vspace{-15pt}
    \label{fig:others2_meanshift_nclusters}
\end{figure}

In order to extract the social circles of the ego networks, the Meanshift algorithm is applied to the annual contact frequency between the ego and its alters, as discussed in Section~\ref{subsection:freq_and_layers}. Figure~\ref{fig:others2_meanshift_nclusters} shows the distribution of the number of clusters found by the Meanshift algorithm, whereby clusters correspond to the social groups around the ego. The figure shows that the most representative cases are those with 4 (the mode) and 5 (the mean) groups, even if a considerable amount of users also have 3 and 6 clusters. This distribution of the optimal number of clusters is also aligned with previous findings in other social networks, e.g., Twitter~\cite{arnaboldi2015online,boldrini2018twitter}.

Table~\ref{tab:n_alters_per_max_clusters} shows the average size of the ego networks as a function of their optimal number of circles. Values similar to Dunbar's model ($\sim$ 150) are found in the case of 6 social circles. On the other hand, egos with a lower number of social circles (i.e., from 3 to 5) show considerably lower values, featuring thus much smaller ego networks. Although these values seem to contradict those displayed in Figure~\ref{fig:others2_meanshift_nclusters} (i.e., the mean of 150 alters), the overall average provided by the figure is the result of combining together all the egos, regardless of their optimal number of circles. Ego with more circles tend to have larger ego networks and they shift the mean toward higher values. However, since the majority of the collected users display 4 and 5 social circles, from both Table~\ref{tab:n_alters_per_max_clusters} and Figure~\ref{fig:others2_meanshift_nclusters}, we conclude that the collected Mastodon sample captures relatively young ego networks, reflecting a likewise young social network, similarly to the early Facebook ego networks from 2009~\cite{arnaboldi2016ego}.

\begin{table}[t!]
    \centering
    %%\vspace{-15pt}
    %\scriptsize
    \caption{Average size of the ego networks as a function of their number of circles}
    %\vspace{-15pt}
    \label{tab:n_alters_per_max_clusters}
    \begin{tabular}{ccc} 
    \toprule
    \textbf{Num.~circles} & \begin{tabular}[c]{@{}c@{}}\textbf{Egos with this}\\\textbf{number of circles}\end{tabular} & \begin{tabular}[c]{@{}c@{}}\textbf{Ego network~size}\\\textbf{(active)}\end{tabular} \\ \midrule
    1            & 72                               & 1.1                        \\
    2            & 67                               & 6.76                       \\
    \textbf{3}   & \textbf{186}                     & \textbf{20.92}             \\
    \textbf{4}   & \textbf{325}                     & \textbf{45.78}             \\
    \textbf{5}   & \textbf{235}                     & \textbf{77.05}             \\
    \textbf{6}   & \textbf{189}                     & \textbf{146.27}            \\
    \bottomrule
    \end{tabular}
\end{table}

\begin{table}[t!]
\begin{minipage}{\columnwidth}
       %\scriptsize
        \caption{Number of alters}
        \label{tab:meanshift_alters}
       \resizebox{\columnwidth}{!}{
        \begin{tabular}{|c|c|c|c|c|c|c|}
        \hline
        & \multicolumn{6}{c|}{\cellcolor{red!30}Circles}\\
         \hline
          \cellcolor{blue!30}Tot circles & \cellcolor{red!15}1 & \cellcolor{red!15}2 & \cellcolor{red!15}3 & \cellcolor{red!15}4 & \cellcolor{red!15}5 & \cellcolor{red!15}6\\
         \hline
         \cellcolor{blue!15}3 & 1.45 &  4.76 & 20.92 & - & - & -\\
         \hline
         \cellcolor{blue!15}4 & 1.42 & 3.74 & 10.61 & 45.78 & - & -\\
         \hline
         \cellcolor{blue!15}5 & 1.17 & 2.82 & 5.78 & 15.36 & 77.03 & -\\
         \hline
         \cellcolor{blue!15}6 & 1.16 & 2.55 & 4.55 & 8.39 & 23.67 & 146.27\\
         \hline
        \end{tabular}
        }
        \vspace{1pt}
       
\end{minipage}

\begin{minipage}{\columnwidth}
        %\centering
        %\scriptsize
        \caption{Scaling ratio}
        \label{tab:meanshift_scaling}
        \resizebox{\columnwidth}{!}{
        \begin{tabular}{|c|c|c|c|c|c|}
        \hline
         & \multicolumn{5}{c|}{\cellcolor{red!30}Circles}\\
         \hline
        \cellcolor{blue!30}Tot circles & \cellcolor{red!15}2/1 & \cellcolor{red!15}3/2 & \cellcolor{red!15}3/4 & \cellcolor{red!15}4/5 & \cellcolor{red!15}6/5\\
         \hline
         \cellcolor{blue!15}3 & 3.38 &  4.39 & - & - & -\\
         \hline
         \cellcolor{blue!15}4 & 2.85 & 2.95 & 4.41 & - & -\\
         \hline
         \cellcolor{blue!15}5 & 2.48 & 2.13 & 2.64 & 4.75 & -\\
         \hline
         \cellcolor{blue!15}6 & 2.27 & 1.81 & 1.85 & 2.78 & 5.91\\
         \hline
        \end{tabular}
        }
        \vspace{1pt}

\end{minipage}
%%\vspace{-25pt}
\end{table}

Table~\ref{tab:meanshift_alters} shows the average number of alters within each circle. When examining all configurations, the values closely approximate those presented in Table~\ref{tab:egonet_offline}, though they tend to be slightly lower. This supports the observation of young but active ego networks. Comparing the configurations suggests that larger ego networks are in a more advanced stage of development. In other words, the outer layers are relatively larger compared to the inner layers, indicating that weaker relationships have already been established, and some of these may evolve into closer connections over time.

It should be noted that, particularly with recently-established platforms, the sizes of the layers may be significantly smaller than expected. However, the scaling ratio among layers often remains a more stable metric, even in these cases~\cite{arnaboldi2012analysis,arnaboldi2015online,arnaboldi2016ego}. Table~\ref{tab:meanshift_scaling} shows the scaling ratio of the sizes between circles. We observe that the scaling ratio is frequently close to 3, aligning with the ``canonical'' scaling ratio in Dunbar's model. A notable exception is the scaling ratio between the last (most external) and second-last layers, which is typically higher in our data. This discrepancy could be due to the relatively short observation window ($\sim$1.5 years) for the considered relationships, making it challenging to characterize relationships in the most external layer, which typically have a contact frequency of about one interaction per year. Despite this, the sizes of the layers and their scaling ratios suggest the existence of the expected ego network structures in Mastodon, even though the social network is still developing.

Table~\ref{tab:meanshift_freq} shows the annual frequency of contact per circle, revealing an interesting pattern. Specifically, when comparing users with different numbers of circles, we observe that the average interaction rate of the most external layers remains remarkably stable across the different groups. As we move from one group to the next (e.g., from users with 3 layers to users with 4 layers), an additional internal layer typically emerges, featuring a higher contact frequency. This, again, suggests that we are observing the initial evolution of user social networks: in this phase, users start by interacting and building layers characterized by lower interaction frequencies (and smaller ego networks overall). As the networks grow, the external layers expand, but some relationships become more intimate, generating additional internal layers with higher interaction frequencies. %COMMENT DAYS PER LAYER HERE
Finally, Table~\ref{tab:meanshift_length} shows the length of relationships in days. It corroborates the observation of Table~\ref{tab:meanshift_freq}, where external layers are almost equivalent in all configurations while additional inner layers appear for more mature ego networks, in this case indicating the emergence of longer and thus more intimate and stable relationships.

\begin{table}[t!]
\begin{minipage}{\columnwidth}
        %\hspace{-10pt}
        %\centering
        %\scriptsize
        \caption{Contact frequency (year)}
        \label{tab:meanshift_freq}
        \resizebox{\columnwidth}{!}{
        \begin{tabular}{|c|c|c|c|c|c|c|}
        \hline
         & \multicolumn{6}{c|}{\cellcolor{red!30}Circles}\\
         \hline
         Max circles\cellcolor{blue!30} & \cellcolor{red!15}1 & \cellcolor{red!15}2 & \cellcolor{red!15}3 & \cellcolor{red!15}4 & \cellcolor{red!15}5 & \cellcolor{red!15}6\\
         \hline
         \cellcolor{blue!15}3 & 35.47 &  20.78 & 7.54 & - & - & -\\
         \hline
         \cellcolor{blue!15}4 & 45.55 & 31.75 & 19.43 & 5.85 & - & -\\
         \hline
         \cellcolor{blue!15}5 & 69.50 & 52.16 & 38.80 & 22.68 & 6.54 & -\\
         \hline
         \cellcolor{blue!15}6 & 84 & 66.19 & 51.71 & 38.54 & 21.95 & 6.46\\
         \hline
        \end{tabular}
        }

\end{minipage}
\begin{minipage}{\columnwidth}
        %%\hspace{-30pt}
        %\centering
        %\scriptsize
        \caption{Bond length (days)}
        \label{tab:meanshift_length}
        \resizebox{\columnwidth}{!}{
        \begin{tabular}{|c|c|c|c|c|c|c|}
        \hline
         & \multicolumn{6}{c|}{\cellcolor{red!30}Circles}\\
         \hline
         \cellcolor{blue!30}Max circles & \cellcolor{red!15}1 & \cellcolor{red!15}2 & \cellcolor{red!15}3 & \cellcolor{red!15}4 & \cellcolor{red!15}5 & \cellcolor{red!15}6\\
         \hline
         \cellcolor{blue!15}3 & 191.47 &  166.26 & 105.37 & - & - & -\\
         \hline
         \cellcolor{blue!15}4 & 208.01 & 186.62 & 159.71 & 96.52 & - & -\\
         \hline
         \cellcolor{blue!15}5 & 229.34 & 210.45 & 190.12 & 103.89 & 97.04 & -\\
         \hline
         \cellcolor{blue!15}6 & 237.99 & 230.08 & 222.99 & 210.40 & 180.57 & 99.51\\
         \hline
        \end{tabular}
        }
\end{minipage}
%\vspace{-30pt}
\end{table}

\hfill

\section{Conclusion}
\label{section:conclusion}

In this study, we investigated user-to-user communication on the Fediverse, focusing on Mastodon as a key platform due to its increased popularity following Twitter's acquisition by Elon Musk. We specifically analyzed users who joined post-acquisition (Others2), representing two-thirds of our sample. These users exhibit a higher tendency for direct communication, particularly amongst themselves. By examining their ego networks, we found evidence supporting Dunbar's social circle model, albeit with smaller circle sizes and a proportionally larger external layer, which are the hallmarks of a ``young'' social network.
These findings indicate that Mastodon is evolving into a platform for cultivating social connections, as users transition from sporadic interactions to closer relationships. This, coupled with the platform's open API and the absence of a suggestion algorithm, makes Mastodon an invaluable resource for studying online communication dynamics, especially given the recent restrictions on Twitter's API. Given these promising results, we plan to replicate this study in the future to assess whether Dunbar's expected circle sizes emerge as the platform matures and whether user interest in Mastodon is sustained. Furthermore, we intend to expand our analysis to other decentralized social networks like Bluesky.

\section*{Acknowledgments}

This work is partially supported by the European Union – Horizon 2020 Program under the scheme “INFRAIA-01-2018-2019 – Integrating Activities for Advanced Communities”, Grant Agreement n.871042, “SoBigData++: European Integrated Infrastructure for Social Mining and Big Data Analytics” (\url{http://www.sobigdata.eu}). This work is also supported by the European Union under the scheme HORIZON-INFRA-2021-DEV-02-01 – Preparatory phase of new ESFRI research infrastructure projects, Grant Agreement n.101079043, “SoBigData RI PPP: SoBigData RI Preparatory Phase Project”. 
This work was partially supported by SoBigData.it. SoBigData.it receives funding from European Union – NextGenerationEU – National Recovery and Resilience Plan (Piano Nazionale di Ripresa e Resilienza, PNRR) – Project: “SoBigData.it – Strengthening the Italian RI for Social Mining and Big Data Analytics” – Prot. IR0000013 – Avviso n. 3264 del 28/12/2021. 
C. Boldrini was also supported by PNRR - M4C2 - Investimento 1.4, Centro Nazionale CN00000013 - "ICSC - National Centre for HPC, Big Data and Quantum Computing" - Spoke 6, funded by the European Commission under the NextGeneration EU programme.
F. Di Cursi, A. Passarella and M. Conti were also supported by the PNRR - M4C2 - Investimento 1.3, Partenariato Esteso PE00000013 - "FAIR", funded by the European Commission under the NextGeneration EU programme.

%\vspace{-10pt}

\bibliographystyle{splncs04}
\bibliography{bibliography.bib}
\end{document}